\documentclass[aip,apl,twocolumn,superscriptaddress]{revtex4}
\usepackage{amsmath,graphicx,dcolumn,bbm}
\newcommand{\ket}[1]{\vert #1 \rangle}
\newcommand{\bra}[1]{\langle #1 \vert}
\begin{document}
\title{Programmable purification of type-I polarization-entanglement}
\author{Simone Cialdi}
\affiliation{INFN, Sezione di Milano, I-20133 Milano, Italia.}
\affiliation{Dipartimento di Fisica, Universit\`a degli Studi di Milano,
I-20133 Milano, Italia.}
\author{Davide Brivio}
\affiliation{Dipartimento di Fisica, Universit\`a degli Studi di Milano,
I-20133 Milano, Italia.}
\author{Matteo G.A.~Paris}
\affiliation{Dipartimento di Fisica, Universit\`a degli Studi di Milano,
I-20133 Milano, Italia.}
\date{\today}
\begin{abstract}
We suggest and demonstrate a scheme to compensate spatial and
spectral decoherence effects in the generation of polarization 
entangled states by type-I parametric downconversion. In our device 
a programmable spatial light modulator imposes a polarization dependent 
phase-shift on different spatial sections of the overall downconversion 
output and this effect is exploited to realize an effective purification 
technique for polarization entanglement.
\end{abstract}
\maketitle
Entanglement is a key resource for several quantum communication and
quantum computation protocols \cite{Nielsen}. Entangled states of
polarization qubit have been realized exploiting spontaneous parametric
down conversion (SPDC) in nonlinear crystals \cite{kwi99} where a pump
photon generates a couple of daughter photons called signal and idler.
Brightness and purity of SPDC-based entanglement sources are limited
by decoherence. Basically, there are two effects that degrade type-I
polarization entanglement. The first is related to the source coherence
time and the second to the emission angle- and frequency-dependent phase
of the SPDC mechanism itself. The effect of a limited coherence of
the source may be reduced using a temporal precompensation technique
\cite{nam02,cia08} whereas entanglement purity is usually recovered by
strong filtering in the angular and spectral SPDC distribution (e.g.
narrow irises and spectral filters). Recently very bright source has
been realized compensating the angular-phase using suitably prepared
birefringent crystals \cite{kwi09}.
\par
In this Letter we suggest and demonstrate a emission-angle and
emission-frequency compensation technique based on the use of a
programmable spatial light modulator ($1$D SLM). This is a crystal
liquid phase mask ($64\times10mm$) divided in $640$ horizontal pixels,
each wide $d=100\mu m$  and with the liquid crystal $10\mu m$ deep. The
SLM is set on the path of signal and idler at a distance $D=500 mm$ from
the two generating crystals. Driven by a
voltage the liquid crystal orientation in correspondence of a certain
pixel changes. The photons with an horizontal polarization feel an
extraordinary index of refraction depending on the orientation, and this
introduce a phase-shift between the horizontal and the vertical
polarizations. Since each pixel is driven independently it is possible
to introduce a phase function dependent on the position on the SLM, i.e.
on the SPDC generation angles. This kind of devices have been already
used as amplitude modulators \cite{lim09}, as well as diffractive 
elements to operate on orbital angular momentum \cite{yao06} and 
for phase-modulation at the
single-photon level \cite{pet08}. The value of our technique is
twofold. On the one hand, our filtering of downconverted photons is
drastically reduced, not only in the angular distribution but also in
frequency. On the other hand, programmability make it easily adjustable
for many implementations.
\par
\begin{figure}[h!]
\includegraphics[angle=270,width=0.75\columnwidth]{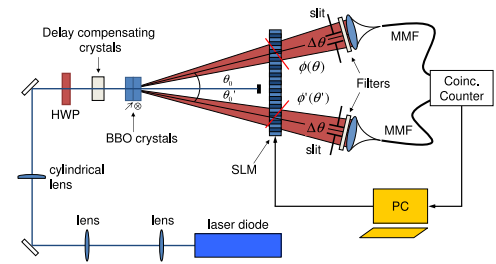}
\caption{(Color online)
Experimental setup. Polarization entangled states are generated
by downconversion in type-I BBO crystals pumped @$405nm$ by a
Newport LQC$405-40$P CW laser diode. Then, a spatial section of
the output cones passes through the SLM which provides
purification of polarization entanglement.
After SLM there are (on both paths) a slit, filters and a lens that
make the image of the pump spot on the crystal into the core of a
multimode optical fiber (MMF).
For tomographic reconstruction we insert a quarter-wave plate,
a half-wave plate and a polarizer and for the optimization of the
phase functions only the polarizers.
}\label{2D_generation}
\end{figure}

In our scheme (see Fig. \ref{2D_generation}) the polarization entangled
state is generated by two adjacent type-I BBO crystals
\cite{kwi99,cia08,cia09}, oriented with their optical axes aligned in
perpendicular planes and pumped by a $405nm$ CW laser diode (Newport LQC$405-40$P).
The part of the pump horizontally polarized generates couples of
vertically polarized photons in the first crystal,
while the vertical pump polarization generates
couples horizontally polarized in the second one. An half-wave plate (HWP)
is used to set the pump polarization at $45^{\circ}$ in order to
balance the probabilities to generate vertical or horizontal couples of photons.
\par
Taking care of the coherent spatial overlap of the two emission
cones the state outgoing the two crystals may be written as
$\ket{\Phi(\omega_p,\omega_s,\theta)}=1/\sqrt{2}
[\ket{HH}+\exp\{\imath \varphi(\omega_{p},\omega_s,\theta)\}\ket{VV}]$,
where $\omega_p$ and $\omega_s$ are the shift from the pump and signal
central frequencies $\omega_p^0$ and $\omega_s^0=\omega_p^0/2$, and
$\theta$ is the shift from the central signal angle $\theta_0$.
Upon expanding all the contributions to the optical paths to first order
we obtain the phase function
$\varphi(\omega_p,\omega_s,\theta)= \phi_0+\alpha L
\omega_p+ \beta L \omega_s - \delta L \theta$, where $L$ is the crystals
length and $\phi_0$ is a constant term depending on the central frequencies
and angles.
The phase term $\alpha L \omega_p$ accounts for the delay
time between horizontal and vertical downconverted photons.
The last two terms appear because the photons generated in the first crystal
must traverse the second one.
These terms may be understood by considering the conservation of the
transverse momentum.
Taking the signal at the fixed angle $\theta_0+\theta$, for different $\omega_s$
the idler sweeps different angles $\theta'_0+\theta'$. This means a
different optical path and explain the phase term $\beta L \omega_s$.  Likewise fixing
$\omega_s$, a positive variation of $\theta$ correspond to a negative
variation of $\theta'$ and this introduces an optical path dependent on
$\theta$, i.e. the phase shift $\delta L \theta$.
The conservation of the transverse momentum implies that the signal and idler
generation angles $\theta_0 +\theta$ and $\theta'_0+\theta'$ are not independent.
We write $\theta'= -\theta +\gamma \omega_s + \gamma' \omega_p$.
Within the pump spectral width the dependence on $\omega_p$ is negligible
and thus we have $\theta'\simeq -\theta + \gamma \omega_s$, with
$\gamma= \partial \theta '/\partial \omega_s=-0.045\times 10^{-15}\,{rad/Hz}=
1.294\times10^{-4}\, {rad/nm}$.
Using this relation and $\delta=2\beta/\gamma$, which can be proved
analytically, the phase function may
be written as $\varphi(\omega_p,\theta,\theta')=\phi_0+
\alpha L\omega_p- \beta/\gamma L \theta+ \beta/\gamma L \theta'$.
We exploit the SLM to add the phase functions $\phi(\theta)$ and $\phi'(\theta')$
to the signal and the idler respectively. This introduces a
polarization dependent shift on the $\ket{HH}$ component of
$\ket{\Phi(\omega_p,\omega_s,\theta)}$ which, together with temporal
delay compensation on the pump, allows one to achieve complete
purification. The overall output state may be written as
\begin{align}
| \Phi \rangle &=\frac{1}{\sqrt{2}} \int_{-\Delta \theta /2}^{\Delta
\theta /2}\!\!\!\!  d\theta \int_{-\Delta \theta /2}^{\Delta \theta /2}
\!\!\!\!d\theta' \int \!\!d\omega_{p}f(\omega_p,\theta_0+\theta,\theta'_0+\theta')
\nonumber \\ &\times  A(\omega_{p})
[e^{\imath\phi(\theta)+\imath\phi'(\theta')}\ket{HH}
+e^{\imath \varphi(\omega_{p},\theta,\theta')}\ket{VV} ]
\label{Eq_Phi}
\end{align}
where $\Delta\theta$ is the angular acceptance due to the presence of
the slit.  $A(\omega_{p})$ is the complex amplitude spectrum of the
pump laser and $f=sinc(1/2 \Delta k_\parallel L)$, where
$\Delta k_{\parallel}=\Delta k_{\parallel}(\omega_p,\omega_s(\theta,
\theta'),\theta)$ is the spectral/angular amplitude of the downconverted
photons \cite{cia09}.
We stress that first order effects, i.e.
frequency- and angle-dependent phase-shifts at the phase mask are
negligible due to the small thickness ($=10\,\mu m$) of the mask.
This definitely show the programmable nature of our device.
\par
\begin{figure}[h!]
\includegraphics[angle=270,width=0.85\columnwidth]{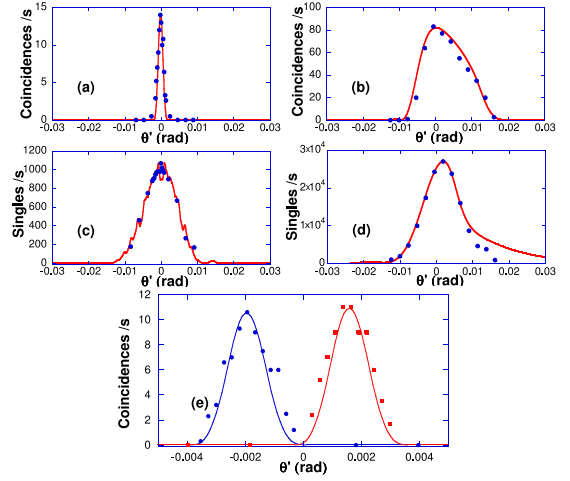}
\caption{(Color online)
Angular distribution of the idler (around $\theta'_0=-3^\circ$)
with signal angle set at $\theta_0+\theta$.  Blue points are the
experimental data; the red solid line the theoretical prediction.
a) Coincidences per second with $F_{BP}$, $\theta=0$; b) Coincidences
per second with $2\times F_{LP}$, $\theta=0$; c) Single counts per second with
$F_{BP}$, $\theta=0$; d) Single counts per second with $2\times F_{LP}$,
$\theta=0$; e)Coincidence per second with $F_{BP}$,
$\theta=\pm1.8mrad$ (blue circle/red square).} \label{ang_distr}
\end{figure}

In order to assess our model and verify the approximation
$\theta'\simeq -\theta +
\gamma \omega_s$ we have measured the angular distribution of the
horizontal component of the idler upon setting the signal at
the fixed angle $\theta_0+\theta$. The angular resolution is
$\Delta\theta\simeq 1.2\,{mrad}$,
corresponding to a slit aperture of $1\,{mm}$. We verify the
absence of diffraction effects using a CCD camera and measure the
angular distribution of coincidences and single counts
using two set of filters set in front of each detector.
We evaluate numerically the distribution of coincidences
$P_{k}(\theta_0,\theta_0') \propto
\int_{\Delta \theta} d\theta \int_{\Delta \theta} d\theta' F_i^2
|f(0,\theta_0+\theta,\theta'_0+\theta')|^2
$
and single counts
$P_{k}(\theta_0,\theta_0') \propto
\int_{\Delta \theta\gg 1} d\theta \int_{\Delta \theta} d\theta' F_i
|f(0,\theta_0+\theta,\theta'_0+\theta')|^2
$($i=1,2$) as obtained omitting integration over pump frequencies
(it can be shown that
$f(\omega_p,\theta_0+\theta,\theta'_0+\theta') \approx
f(0,\theta_0+\theta,\theta'_0+\theta')$
for crystals length $L\lesssim 1 mm$)
and using
$F_1=F_{BP}$, a $10\,nm$ bandpass filters
and $F_2=F_{LP}^2\times QE$, i.e. a filter built by two $F_{LP}$
longpass  filters (cut-on wavelenght $715\,nm$) and the quantum efficiency
$QE$ of each detector (taken from the specification
of the photodiode $C30921S$).
In Fig. \ref{ang_distr} we compare the experimental results (points)
with the theoretical distributions (solid lines): for panels (a,b,c,d)
$\theta_0=3^\circ$, whereas for panel (e) the signal is set at
$\theta_0\pm\theta$, with $\theta=1.8\,mrad$. In Fig.
\ref{ang_distr} (a) and (e) the experimental data are deconvolved
with  the function of the slit \cite{ria86}. As it is apparent from the
plot we have very good agreement with the theoretical expectations. The
mismatch in the right part of panel (d) is due to the finite
size of the phase mask.
\par
For the purification procedure we set
the slits at $\Delta \theta\simeq 6.5 mrad$ and use, on both channel,
the two filters $F_{LP}$. This allows us to collect the downconverted
photons within a wide spectrum and angular distribution. To collect as
many photons as possible we make the imaging of the pump spot on the
crystals ($\simeq 1.5\,{mm}$) into the optical fibers core (diameter of
$62.5\mu m$) using the coupler lenses; the
coincidence counts are about $100/s$, basically due to low quantum 
efficiencies of our home made detectors ($<10\%$). Notice that the 
angular acceptance $\Delta \theta$ acts as a $100nm$ bandpass
spectral filter for the down converted photons.  The delay time between
the photons may be compensated upon the introduction of a proper
combination of birefringent crystals on the pump path, as already
demonstrated in \cite{nam02,cia08} (see Fig. \ref{2D_generation}).
Upon setting $\phi'(\theta')= \beta/\gamma L \theta' + \epsilon'$ and
$\phi(\theta)=-\beta/\gamma L \theta + \epsilon$, with
$\epsilon'+\epsilon=\phi_0$, it is possible to compensate all the
remaining phase terms in $\varphi (\omega_{p},\theta,\theta')$ and to
achieve complete purification of the state. With this choice we may generate
polarization entangled state of the form (\ref{Eq_Phi})
without any phase term. It is worth noting that
the SLM also replaces the birefringent plate used for the optimal
generation of photon pairs \cite{kwi99}. Using the SLM we set the
phase functions $\phi(x)=a_2(x-x_{c2}) +b_2$, and
$\phi'(x)=a_1(x-x_{c1})+ b_1$, where $x$ is the pixel number,
$x-x_{c2}=\frac{D}{d}\theta$ and $x-x_{c1}=\frac{D}{d}\theta'$, $x_{c1}$
and $x_{c2}$ are the central pixels on idler and signal, i.e. the pixels
corresponding to the central angles $\theta'_0$ and $\theta_0$. The
values of the parameters $a_1, b_1, a_2$ and $b_2$ has been optimized
upon inserting two polarizers set at $\alpha_1=45^{\circ}$ and
$\alpha_2= -45^{\circ}$ in front of the couplers and then searching for
the minima in the coincidence counts, corresponding to the values of
$b_{1,2}$ compensating the constant phase difference $\phi_0$ and
$a_{1,2}$ removing the angular dependence on $\theta$ and $\theta'$. For
our configuration $a_1=-a_2=\beta L d/\gamma D\simeq - 0.05$,
$b_1+b_2=\phi_0$.
\begin{figure}[h]
\includegraphics[angle=270,width=0.85\columnwidth]{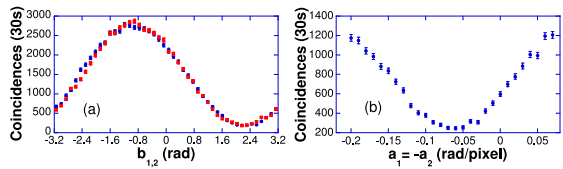}
\caption{(Color online)
Coincidence counts on a time window of $30s$ with
the polarizers in front of the couplers set at $45^{\circ}$ and
$-45^{\circ}$. (a) Coincidences as a function of $b_{1,2}$ (blue/red)
with $b_{2,1}=0$ and optimal $a_{1,2}$; (b) Coincidences as a function
of $a_1=-a_2$ with optimal $b_1$ and $b_2=0$.
\label{scan_a_and_b} }
\end{figure}
\par
In Fig.  \ref{scan_a_and_b}(a) we report the coincidence counts on a
time window equal to $30s$ as a function of $b_{1}$ ($b_2$) (blue/red)
with $b_2=0$ ($b_1=0$) and with $a_{1,2}$ set to their optimal values.
In Fig. \ref{scan_a_and_b}(b) we report the coincidence counts on a time
window of $30s$ as a function of $a_1=-a_2$ with $b_2=0$ and
$b_1=\phi_0$. The agreement with the theoretical model is excellent.
In turn, the purification of the state works as follows: starting from a
visibility equal to $0.423\pm 0.016$ we achieve $0.616 \pm 0.012$ after
the delay compensation with the crystals and $0.886 \pm 0.012$ after the
spatial compensation with the SLM. The residual lack of visibility is
due to low spatial coherence of the pump, which is spatially multimode.
Upon adding a constant phase $\zeta$ to the term $b_1$ or $b_2$ we
may also generate a wide zoology of entangled states. With $\zeta=0,\pi$
we have, in ideal conditions, the Bell states $\ket{\Phi^{\pm}}$
whereas with $\zeta=\pi/2$ we  may generate the state $\ket{\Phi_{\pi/2}}
=1/\sqrt{2}[\ket{HH}+\imath \ket{VV}]$.
\par
\begin{figure}[h!]
\includegraphics[width=0.8\columnwidth]{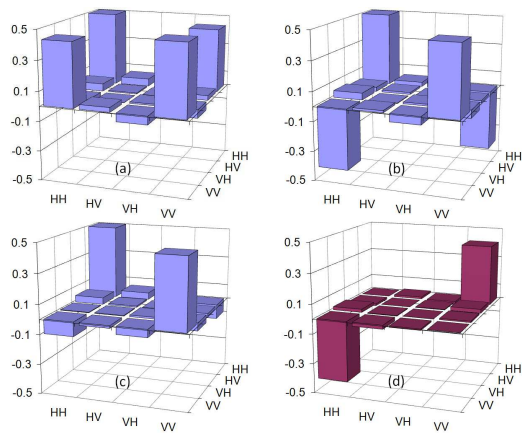}
\caption{(Color online)
Characterization of the output state. We report the tomographic
reconstruction of the purified polarization entangled state
(a)$\ket{\Phi^+}\bra{\Phi^+}$ (real part), (b)
$\ket{\Phi^-}\bra{\Phi^-}$ (real part), (c) real and (d) imaginary part
of $\ket{\Phi_{\pi/2}}\bra{\Phi_{\pi/2}}$.}\label{qt} \end{figure}

In order to characterize the output state and to check the effects of
the decoherence processes, we have performed state reconstruction by
polarization qubit tomography. The procedure goes as follows: we
measure a suitable set of independent two-qubit projectors
\cite{mlk00,jam01} corresponding to different combinations of polarizers
and phase-shifters and then the density matrix is reconstructed from the
experimental probabilities using maximum-likelihood reconstruction of
two-qubit states. The tomographic measurements are obtained by
inserting a quarter-wave plate, a half-wave plate and a polarizer.
The results are summarized in
Fig.  \ref{qt}. Fidelities of the purified polarization states
$\ket{\Phi} =\ket{\Phi^+}$, $\ket{\Phi^-}$ and $\ket{\Phi_{\pi/2}}$ are
$F\simeq 0.90 \pm 0.01$, where uncertainty has been evaluated by
propagating the errors on the determination of matrix elements by
tomography. In order to achieve this precision we have employed a long
acquisition time ($\sim 60 s$) thus also demonstrating the overall
stability of our scheme. The limited value of the fidelities is basically 
due to spatial incoherence of the pump and the incomplete temporal 
delay compensation.
\par
In conclusions, we have suggested and implemented a scheme to
compensate spatial and spectral decoherence effects of type-I SPDC-based
entangled states. In our device a programmable spatial light modulator
acts on different spatial sections of the overall downconversion output
and provides polarization entanglement purification. Our device allows
the effective generation and purification of type-I polarization entanglement
and paves way for quantum information processing with low-cost,
low-coherence, low-power source of entanglement \cite{cia10}.

\end{document}